\newcommand{\Tr}{\rm Tr \,}
\newcommand{\tr}{\rm tr \,}
\newcommand{\pslash}{\overlay {\slash }{p}}
\documentstyle[prd,aps,preprint]{revtex}
\begin{document}

\preprint{ECT*/Dec/95-003}

\draft

\title{Dynamical Higgs Mechanism without Elementary Scalars: 
\\ A Lesson from Instantons }

\author{Dmitri Diakonov}

\address{Petersburg Nuclear Physics Institute,
         Gatchina, St. Petersburg 188350, Russia}

\author{Hilmar Forkel {\it and}  Matthias Lutz }

\address{ECT*, Villa Tambosi,
         I-38050 Villazzano, Italy}

\date{November, 1995}
\maketitle

\begin{abstract}

The generation of gauge--dependent fermion vacuum condensates in 
Yang--Mills theory would dynamically break the gauge symmetry and 
thus provide an alternative to the Higgs mechanism in unified 
theories. We explore a simple example: instanton-induced quark 
pair (diquark) condensation in QCD with 2 flavors and $N_c$ colors. 
We do find diquark condensates in the vacuum, but only for $N_c = 2$, 
where they are equivalent to the standard quark-antiquark condensate. 
At $N_c \geq 3$ diquark condensates exist as meta--stable saddle 
points of the instanton-induced effective QCD action and may strongly 
affect the properties of matter under extreme conditions. The scalar 
diquark excitation, which is a Goldstone boson at $N_c =2$, becomes 
unbound for $N_c \geq 3$. 

\end{abstract}



Instanton-induced quark interactions provide a well-known mechanism 
for spontaneous chiral symmetry breaking by quark-antiquark pair 
condensation in the QCD vacuum \cite{dia95}. In this letter we persue 
the question whether these interactions, which are attractive in certain 
quark-quark channels, might also induce diquark condensates. Since 
such condensates are not gauge invariant (except for $N_c =2$), they
would signal dynamical gauge symmetry breaking. 

Diquark condensates, which are a direct analogue of Cooper pair 
condensates in superconductive materials, could furthermore 
serve as a prototype for dynamical gauge symmetry breaking in other 
theories and thus as an alternative to the standard Higgs mechanism. 
Even if they are absent in the ground state of a theory, however, 
they might still form in a state corresponding to a saddle point of 
the effective action. Such meta-stable states could, for example, 
play an important role in cosmology and in the evolution of the 
universe.

Our study is based on the effective quark interaction 
\cite{dia95,thooft}
\begin{eqnarray}
{\cal L} & = & g \left\{
\left(1- \frac{1}{N_c} \right) \left[ (\bar{q} q)^2 + 
(\bar{q} \gamma_5 q)^2 - (\bar{q} \vec{\tau} q)^2 - 
(\bar{q} \gamma_5  \vec{\tau} q)^2 \right] \right. \nonumber \\
&-& \left. \frac{1}{N_c} \left[(\bar{q} C \vec{\lambda}_A i \tau_2 
\bar{q}^{ T}) (q^T C \vec{\lambda}_A i \tau_2 q)  + 
(\bar{q} \gamma_5 C \vec{\lambda}_A i \tau_2 \bar{q}^{ T} )
 (q^T \gamma_5 C \vec{\lambda}_A i \tau_2 q) \right] \right\}\, ,
\label{lagr}
\end{eqnarray}
which is induced by the fermionic zero-modes in the instanton 
background (in euclidean space-time). Here, $\vec{\tau}$ are the $SU(2)$ 
flavor Pauli-matrices, the $\vec{\lambda}_A$ form the antisymmetric 
subset of the $SU(N_c)$ color generators and $C$ is the Dirac charge 
conjugation matrix. 

In order to probe the condensation of quark-quark and quark-antiquark 
pairs in the vacuum, one can bosonize the interaction (\ref{lagr}) by
introducing auxilary scalar, pseudoscalar, scalar diquark and 
anti-diquark fields and integrating the quarks out \cite{diqbos}. 
The resulting bosonic effective action can then be treated in leading 
loop approximtion, i.e. by neglecting meson and diquark loops. Instead, 
we will use an equivalent \cite{Praschifka}, but technically more 
efficient approach. It is based on a somewhat more general, bilocal 
effective action functional $\Gamma[\Delta,D,\bar{D}] 
$\cite{Praschifka,CJT}, which determines in our case the ground state 
energy of the dynamics (\ref{lagr}) under the constraint that the 
vacuum expectation values of three time-ordered quark bilinears,
\begin{eqnarray}
\Delta(x) = <0|T\, q (x) \bar{ q } (0)|0> \; ,\\
D(x) = <0|T\, q (x) q (0)|0> \;, \label{ddef} \\ 
\bar{D}(x) = <0|T\, \bar{q} (x) \bar{q} (0)|0> \;. \label{dbardef} 
\end{eqnarray}
are fixed. In terms of $\Delta, D$ and $\bar{D}$, the effective action 
can be expressed as \cite{Praschifka,CJT}
\begin{eqnarray}
\Gamma [\Delta,D] = &-& \frac12 \Tr \log
\left(  \begin{array}{lr} \bar{ D} & - \Delta^{T} \\
\Delta & D \end{array} \right) \nonumber \\
&+& \Tr \left( \Delta_0^{-1} \Delta - 1 \right)
+ V[\Delta,D,\bar{D}] , \label{effa}
\end{eqnarray}
which consists of a kinetic term, containing the free quark propagator 
$\Delta_0 (p) = i (\pslash -m_0+ i \epsilon)^{-1}$ with the current quark 
mass $m_0$, and the sum of all two-particle-irreducible vacuum diagrams 
$V[\Delta,D,\bar{D}]$. (The symbol $\Tr$ denotes a trace over color, 
flavor and Dirac indices as well as over the space-time dependence.) 

The physical propagator $\Delta $ and $D, \bar{D}$ are obtained as 
the stationary point of $\Gamma $ with minimal energy. It is 
straightforward to show \cite{help} that all stationary points of the 
effective action correspond to solutions\footnote{Note that the  
{\it solutions} of the gap equations do not necessarily satisfy the 
condition $\gamma_0 \bar D \gamma_0 = D^{\dagger } $ (see, e.g., eqs. 
(\ref{sol3} -- \ref{sol2}) which one would naively expect from 
(\ref{ddef}), (\ref{dbardef}) and $\bar q = q^{\dagger } \gamma_0$. 
In the effective 
action functional, $\Gamma$, however, $D$ and $ \bar D$ are 
independent quantities, and the constraint from the operator identity 
$\bar{q} = q^{\dagger } \gamma_0$ is not mandatory. In the path
integral formulation the invariance of the measure under an 
infinitesimal transformation $ q \rightarrow q + i \delta \alpha \,q$ 
and $ \bar{q} \rightarrow \bar{q} - i \delta \alpha \, \bar{q} $ leads 
to a less stringent constraint, $ \Tr D\,\frac {\delta \Gamma}{\delta D} 
=\Tr \bar D\,\frac {\delta \Gamma}{\delta \bar D}$ (consistent with 
(\ref{sol3} -- \ref{sol2})).} of the coupled set of gap equations
\begin{eqnarray}
\frac{\delta V}{\delta \Delta} 
&=& \left( \Delta + D (\Delta^T)^{-1} \bar{D} \right)^{-1} 
    - \Delta_0^{-1} \label{gap1} \\
\frac{\delta V}{\delta D} &=& \frac{1}{2} \Delta^{T-1} \bar{D}
       \left( \Delta + D (\Delta^{T})^{ -1} \bar{D} \right)^{-1} \, ,\\ 
\frac {\delta V} {\delta \bar{D}} &=& \frac{1}{2} 
      \left( \Delta + D (\Delta^{T})^{-1} \bar{D} \right)^{-1} 
      D (\Delta^T)^{-1} \, . \label{gap2}
\end{eqnarray}
We now specify the potential $V$ by adopting the Bogoliubov-Hartree 
approximation \cite{Kleinert}, which is exactly equivalent to the 
zero-meson and -diquark loop approximation in the bosonization 
language \cite{Praschifka}. Restricted to the terms which can affect  
scalar quark pair and quark-antiquark condensation, $V$ reads
\begin{eqnarray}
V[\Delta, D,\bar{D}] &=& i\, g V_4 \frac { N_c-1 }{N_c} 
  \left[ (\tr \Delta(0) ) \; (\tr \Delta(0) ) 
  - (\tr \vec{\tau } \Delta(0)  ) \; (\tr \vec{\tau } \Delta(0)) \right] 
\nonumber\\
  &+&i\, \frac {g V_4} {2 N_c}  (\tr \tau_2 \vec{\lambda}_A \gamma_5  
C \;D^{\dagger }(0) )  ( \tr \tau_2 \vec{\lambda}_A \gamma_5 C^{-1} D(0) )
\nonumber\\
  &+&i\, \frac {g V_4} {2 N_c}  (\tr \tau_2 \vec{\lambda}_A \gamma_5  
C \;\bar{D}(0) )  ( \tr \tau_2 \vec{\lambda}_A \gamma_5 C^{-1} 
\bar{D}^{\dagger }(0) ),       
\end{eqnarray}
where $V_4 \equiv \int d^4x$ is the spacetime volume and the trace 
($\tr$) is over color, flavor and Dirac indices only.

Under the assumption of a flavor-symmetric vacuum and with $N_c = 2$ or 
$3$, the most general ansatz\footnote{We preselect the standard chiral 
orientation of the quark-antiquark condensate (i.e. $< \bar{q} \gamma_5 
q > = 0$) and do not consider potentially interesting colored 
quark-antiquark condensates without simultaneous quark 
pair condensation in the present paper.} for $D$ and $\bar{D}$
yields solutions to the gap equations of the form
\begin{eqnarray}
\Delta (p) &=& \frac {i} {\pslash -m + i \epsilon} 
             \left(
             1 + (\vec \alpha \cdot \vec \lambda_A)^2 \frac {m_c^2}{p^2-m^2-m_c^2+i \epsilon}
             \right),  \label{sol1} \\
D(p) &=& \tau_2 \, \vec \alpha \cdot \vec \lambda_A \,C \gamma_5 
\frac { i \, m_c}{p^2-m^2-m_c^2+i \epsilon}, \label{sol3}\\
\bar{D}(p) &=& \tau_2 \, \vec \alpha \cdot \vec \lambda_A \, 
\gamma_5 C^{-1} 
\frac { i \, m_c}{p^2-m^2-m_c^2+i \epsilon},
\label{sol2} 
\end{eqnarray}
which are expressed in terms of the constituent quark mass $m$, 
an analogous mass parameter, $m_c$, related to the quark pair 
condensation  strength, and the $N_c (N_c-1)/2$ dimensional unit vector 
$\vec \alpha $ (i.e. $ \vec \alpha \cdot \vec \alpha =1$), which 
determines the orientation of a potential diquark condensate in the 
$so(N_c)$ subalgebra of color space. Since only for $N_c =2$ and $3$ 
the expression $(\vec \alpha \cdot 
\vec \lambda_A)^2 $ is a projection operator\footnote{
For all $N_c$,  $A^4 = \frac{1}{N_c} \tr \it (A^2) A^2 + (\vec \beta \cdot 
\vec \lambda_A) A$, where $A \equiv (\vec \alpha \cdot \vec \lambda_A)$ 
and $\beta_i = \alpha_j \, \alpha_k \, \alpha_l \, d_{i j m} \, d_{k l m}$ 
in terms of the totally symmetric d-symbols of the $su(N_c)$ Gell-Mann 
matrices. Therefore, only for $N_c = 2$ with $\beta_i = 0$ and for $N_c 
= 3$ with $\beta_i = \frac16 \alpha_i \tr (A^2)$, does $A^2$ become a 
projector for all $ \vec \alpha $ of unit length.} with $(\vec \alpha 
\cdot \vec \lambda_A)^4 = (\vec \alpha \cdot \vec \lambda_A)^2$, and 
since this property is necessary to satisfy the color structure of the 
gap equations, the above solutions do not generalize to $N_c >3$ for 
general $\vec \alpha$. Special choices of $\vec \alpha$, however, 
as for example $\vec \alpha = \{1,0,0,...\}$, lead to solutions 
for all $N_c$.

Inserting these solutions into (\ref{effa}), 
we obtain their energy density $\epsilon  = i \Gamma /V_4$:
\begin{eqnarray}
\epsilon[m^2,m_c^2] &=&- \left( N_c-2 \right) \epsilon_0[m^2] 
          - 2 \epsilon_0[m^2+m_c^2] \nonumber \\
      &+&  (N_c -2) m (m-m_0) I[m^2] + 2 (m (m-m_0)+m_c^2) I[m^2+m_c^2] \nonumber\\
      &-& g \frac {N_c-1}{N_c} 
               \left( (N_c-2) m I[m^2] + 2\, m I[m^2+m_c^2] \right)^2 
   \nonumber\\
      &-& g \frac {1 } {N_c}  \left(2\, m_c I[m^2 + m_c^2] \right)^2, 
\label{edens}
\end{eqnarray}
where the integrals $\epsilon_0[m^2]$ and $ I[m^2] $, properly 
regularized \cite{lut92} by the cutoff $\lambda $, are given by
\begin{eqnarray}
\epsilon_0[m^2] &=& 4\,i \int \frac { d^4 \, p}{(2 \pi )^4} 
\log [p^2-m^2+i\, \epsilon ] \nonumber \\
&=& \frac {2}{\pi^2} \int_0^{\lambda } d \,p \;p^2 \sqrt{p^2+m^2}\\
I[m^2] &=&  2\, \frac {\partial \epsilon_0}{\partial m^2} .
\end{eqnarray}
The gap equations in terms of the variables $m$ and $m_c$ can now be 
recovered from the stationarity requirement for the energy density:
\begin{eqnarray}
m-m_0 &=& g \frac {N_c-1}{N_c}  
      \left[
      2 (N_c -2) \,m I[m^2] + 4\,m I[m^2+m_c^2] \label{altgap1}
      \right], \\
m_c &=& g \frac {1 }{N_c} 4\,m_c I[m^2 + m_c^2]. \label{altgap2}
\end{eqnarray}

Following reference \cite{dia95} the coupling constant $g$ can be 
related to the gluon condensate $<F^2_{\mu \nu}> = 32 \pi^2 \bar{n}$ or 
the instanton density $ \bar{n} $ by adding to the vacuum energy 
the term
\begin{equation}
\Delta \epsilon =2 \bar{n} \log g
\end{equation}
and minimizing the resulting expression with respect to the coupling 
strength $g$. This leads to a further equation,
\begin{eqnarray}
\frac{2 \bar{n}}{g} &=& \frac {N_c-1}{N_c} 
                  \left( (N_c-2) m I[m^2] + 2\, m I[m^2+m_c^2] \right)^2 
\nonumber\\
                &+& \frac {1 } {N_c}  \left( 2\,m_c I[m^2 + m_c^2] 
\right)^2 \; .\label{couplfix}
\end{eqnarray}
We fix the instanton density at the value $\bar{n} =$ $N_c$/3 (229 
MeV)$^4$ \cite{dia95}, which incorporates a straightforward extrapolation 
of the $N_c$ dependence for $N_c \neq 3$. The cutoff is set to $\lambda = 
640 $ MeV, which leads to the value $<\bar{q} q> = - 251{\rm MeV} $ 
of the quark condensate for $N_c =3$, consistent with phenomenology.

The explicit expressions (\ref{edens}) for the energy density and (\ref{altgap1}), (\ref{altgap2}) for the gap equations show a pronounced 
$N_c$ dependence. We will thus discuss the three distinct cases $N_c=2$, 
$N_c=3$ and $N_c >3$ separately. 

a) $N_c=2$:

In the chiral limit ($m_0 =0 $), this is the unique case in which 
the energy density $\epsilon = \epsilon [m^2+m_c^2]$ depends solely on 
the combination $m^2+m_c^2$, and where the gap equations are consequently 
symmetric under exchange of $m$ and $m_c$. Both of these properties 
reflect the $SU(4)$ (Pauli-G{\"u}rsey \cite{PG}) symmetry of the 
instanton-induced dynamics \cite{dimitri} and of QCD for two colors 
\cite{pes80}, which mixes quark and antiquark states and leaves 
$m^2+m_c^2 $ invariant. If the coupling $g$ 
becomes strong enough to generate a quark-antiquark condensate, the 
latter can be transformed into an equivalent, degenerate quark-quark 
pair condensate by an $SU(4)$ transformation. 

For the same reason, such a diquark condensate does not break any more  symmetries than the usual $< \bar{q} q>$ condensate, and in particular 
neither baryon number nor color symmetry. Indeed, these diquark 
condensates have the quantum numbers of colorless  $N_c = 2$ ``baryons'' 
and are thus gauge invariant. Furthermore, should one decide to 
quantize the theory in a vacuum sector containing condensed diquarks, 
then the correponding $SU(4)$ transformation of the baryon 
charge generator would reveal that it also has zero baryon number.

With the coupling $g$ fixed from the instanton dynamics via eq. 
(\ref{couplfix}), the energy density $\epsilon$ has the form of a 
mexican hat, and thus the symmetry breaking solution with $m^2+m_c^2=  
(271 {\rm MeV})^2$ is energetically favored compared to the symmetry 
preserving one, $m=m_c=0$. The energy density is plotted in Fig.1a. 

The above discussion implies that the diquark vacuum condensate is 
not at odds with the Vafa-Witten theorem \cite{vaf84}, which states 
that vector symmetries cannot be spontaneously broken in vector-like 
gauge theories. Since this theorem has been established only for 
finite current quark masses, however, we also considered the case $m_0 
\neq 0$, which breaks the $SU(4)$ symmetry explicitly and 
thereby lifts the degeneracy of the vacuum in favor of a unique ground 
state. For $m_0 = 11 $ MeV, the resulting energy density is shown in Fig. 
1b. Indeed, the mexican hat potential is tilted such that the absolute 
minimum occurs at $m_c =0 $ and $m = 346 $ MeV, which is in accord with 
the Vafa-Witten theorem. 

b) $N_c=3 $

In the (physical) case of three colors, the energy density is shown in 
Fig. 2a for $m_0 =0$. Its absolute minimum corresponds to a constituent 
quark mass of $m=346$ MeV and a vanishing quark pair parameter $m_c=0$. 
This implies (see ref. \cite{dia86}) $<\bar{q} q> =$ - (251 
{\rm MeV})$^3$ for the quark condensate, a vanishing diquark condensate, 
and $f_{\pi } = 95$ MeV for the pion decay constant. 

Fig.2a also shows a saddle point of $\epsilon $ at $m=0$ MeV and $m_c= 
449$ MeV, corresponding to a pure diquark condensate state. Its energy 
density is $ (282 {\rm MeV})^4$ larger than that of the vacuum solution.
The symmetry preserving local maximum with $m=m_c=0$ lies $ (629
{\rm MeV})^4$ above the vacuum. 

Such saddle points can acquire physical significance in the presence 
of large temperatures or baryon densities, e.g. as a transition state 
for baryon number or color symmetry violating processes. They could 
thus play a role in the evolution of the early universe, inside stars 
and in ultrarelativistic heavy-ion collisions. These interesting 
possibilities deserve further study. 

Fig. 2b shows the energy density for a finite current quark mass of 
$m_0 =11$ MeV. The absolute minimum is shifted to $m=353$ MeV with 
$m_c=0$, whereas the meta--stable saddle point moves to $m_c=451 
\, {\rm MeV}$ and $m=-4 \, {\rm 
MeV}$. Thus the constituent mass of quarks propagating in this state 
becomes slightly negative for finite $m_0$, which can be directly 
understood from the $m_0$-dependent terms in the energy density 
(\ref{edens}). The latter are repulsive for negative constituent 
masses, which are prefered since the saddle point is a maximum in 
the $m$ direction. Of course negative quark masses do not render 
the theory tachyonic and occur, for example, in QCD with a finite 
$\theta$ vacuum angle.

c) $N_c>3$:  

As mentioned above (below eq. (\ref{sol3})), our quantitative results 
for $N_c >3$ are based on a somewhat more restrictive ansatz for the 
color structure of the quark and diquark propagators than that for 
$N_c = 2,3$. Besides the trivial solutions with $m=0$ and $m_c=0$, one 
still finds symmetry breaking solutions (for sufficiently large, 
attractive $g$) with $m>0$ and $m_c=0$ or with $m=0$ but $m_c>0$. 
However, it is easily established that the gap equations (\ref{altgap1}) 
and (\ref{altgap2}) do not support solutions with both nonzero $m$ 
and $m_c$. 

Estimating the coupling as before from eq. (\ref{couplfix}),
moreover, we find the same qualitative condensation pattern 
as for $N_c = 3$: The vacuum contains solely a quark-antiquark 
condensate, whereas a state with condensed quark pairs exists as 
a saddle point of the effective action. The energy difference 
between these two states increases with $N_c$, as expected from 
eq. (\ref{lagr}): for large $N_c$ the attraction in the diquark 
channel is suppressed relative to that in the quark-antiquark 
channels, and at $N_c \rightarrow \infty$ diquark degrees of 
freedom can be neglected altogether.

We checked that an artificial enhancement of the attraction in 
the diquark channel relative to that in the quark-antiquark 
channel\footnote{Note 
that this is an ad-hoc modification of the instanton-generated 
interaction, which leaves the realm of $N_c$-QCD.} generates a 
diquark condensate also in the ground state. Inspection of the 
$m_0$-dependent terms in the energy density (\ref{edens} reveals, 
furthermore, that the diquark vacuum condensate appears at $m 
\geq 0$. This is in contrast to the saddle point solutions 
discussed above. 

A diquark--condensed vacuum phase such as the one generated above 
(or also colored $\bar{q} q$ condensates) could be of interest as a 
prototype for dynamical gauge symmetry 
breaking by instantons. In view of possible applications to unified 
theories, one should keep in mind that the Vafa-Witten theorem does 
not exclude diquark vacuum condensates in the presence of 
axial-vector gauge couplings, as they occur in the weak interaction 
sector.  

We finally turn to the calculation of the mass of the scalar diquark 
excitation on top of the ordinary vacua (with $g$ fixed by eq. 
(\ref{couplfix})). This mass can be readily obtained by 
solving the Bethe-Salpeter equation with the appropriate kernel given 
by $K_{BS} = \delta^2 V/ (\delta D \delta \bar{D }) $. The mass 
equation reads:
\begin{equation}
J[m_s^2] = \frac{N_c}{ 2g} 
         = N_c \left( N_c-1 \right) I[m^2] 
\end{equation}
with 
\begin{equation}
J[q^2] = 2 \left( 2
             + \sum_{n=0}^{\infty } \frac {(-1)^{n+1}} {(2 n+1)!!}
              \left( \frac {q^2}{2} \frac {\partial}{\partial m^2}
           \right)^{n+1}
           \right) \frac {\partial \epsilon_0}{\partial m^2} 
\end{equation}
Identifying the quark condensate $<\bar{q} q>$ and the square of the 
pion decay constant $f_{\pi }^2 $ with appropriate derivatives of the 
free vacuum energy $\epsilon_0 $ as given below,
\begin{eqnarray}
<\bar{q} q> &=& -N_c m  \frac {\partial}{\partial m^2} \epsilon_0[m^2],\\
f_{\pi }^2 &=& - N_c m^2
      \left( \frac {\partial }{\partial m^2} \right)^2 \epsilon_0[m^2],
\end{eqnarray}
we can write the scalar diquark mass as  
\begin{equation}
m_s^2 = - 2 \frac {<\bar{q} q> m}{f_{\pi }^2} \left( N_c (N_c-1) -2 \right)
 + O [m_s^4]
\end{equation}
We observe that for $N_c=2$ we recover $m_s=0 $; i.e. the scalar 
diquark and the pion are the Goldstone bosons of the spontaneously 
broken $SU(4)$ symmetry. However, already for $N_c=3$ the scalar 
diquark mass is pushed above the quark-quark threshold, up to about 
$1$ GeV. The instanton sector of QCD does therefore not support the 
existence of a light scalar diquark for $N_c =3$. We consider this 
result an important by-product of our study, since the possibility 
of a light diquark is widely discussed in the current literature 
\cite{ans93}.

To summarize, we studied instanton-induced quark 
interactions in $SU(N_c)$ QCD with self-consistently fixed couplings. 
In mean-field approximation (i.e. neglecting meson and diquark loops), 
these interactions do not generate quark-pair condensates in the 
vacuum state, except for the special case of two colors, where the 
additional $SU(4)$ symmetry renders them equivalent to the usual 
quark-antiquark condensate. We also do not find light scalar diquark 
states for $N_c \geq 3$. Since quark-pair condensates for $N_c >2$ 
would break color and further discrete and continuous symmetries, 
our results are in accord with phenomenolgy for QCD with $N_c=3$ 
and with general expectations for $N_c=2$ and $N_c >3$.
 
In other dynamical settings, however, as for example in grand unified 
theories, instanton-induced interactions could induce vacuum condensates 
which break the gauge symmetry of the interactions down to a subgroup.
Such a mechanism could furnish a viable alternative to the standard 
Higgs sector of unified theories, which is of dynamical origin and 
does not require the presence of scalar fields. A prototype of such 
a situation can also be generated in a modified version of our dynamical 
framework, where the coupling in the diquark channel is increased 
beyond a critical value. 

Even in the case of QCD with $N_c$ colors (i.e. with the effective 
coupling self-consitently fixed by the instanton sector), however, 
quark pair condensates do exist in metastable states, corresponding 
to saddle points of the energy density. It is tempting to speculate 
that such states could have played a role in earlier phases of the 
evolution of the universe or that they could be found inside of hot 
stars or in ultra-relativistic heavy-ion collisions. 

D.D. acknowledges the hospitality of the ECT* in Trento, where most 
of this work was done, and partial support under grant INTAS-93-1201.
H.F. and M.L. acknowledge support by the European Community through 
the HCM programme.

\newpage

\begin{figure}
\caption{The energy density $\epsilon$ at $N_c =2$ for a) $m_0=0$ 
(i.e. in the chiral limit) and b) $m_0=11 \, {\rm MeV}$.} 
\label{fig1}
\end{figure}

\begin{figure}
\caption{The energy density $\epsilon$ at $N_c =3$ for a) $m_0=0$ 
(i.e. in the chiral limit) and b) $m_0=11 \, {\rm MeV}$.} 
\label{fig2}
\end{figure}


\begin{references}
\bibitem{help} For this purpose, the following identity is useful:
  \begin{eqnarray}
   \Tr \log
   \left(  \begin{array}{clcr}
               \bar{D} & - \Delta^{T} \\
                \Delta & \quad D \end{array} \right)
    = \Tr \log \Delta 
    + \Tr \log \left( \Delta + D (\Delta^T)^{-1} \bar{D} \right) 
\nonumber 
   \end{eqnarray}
\end{references}
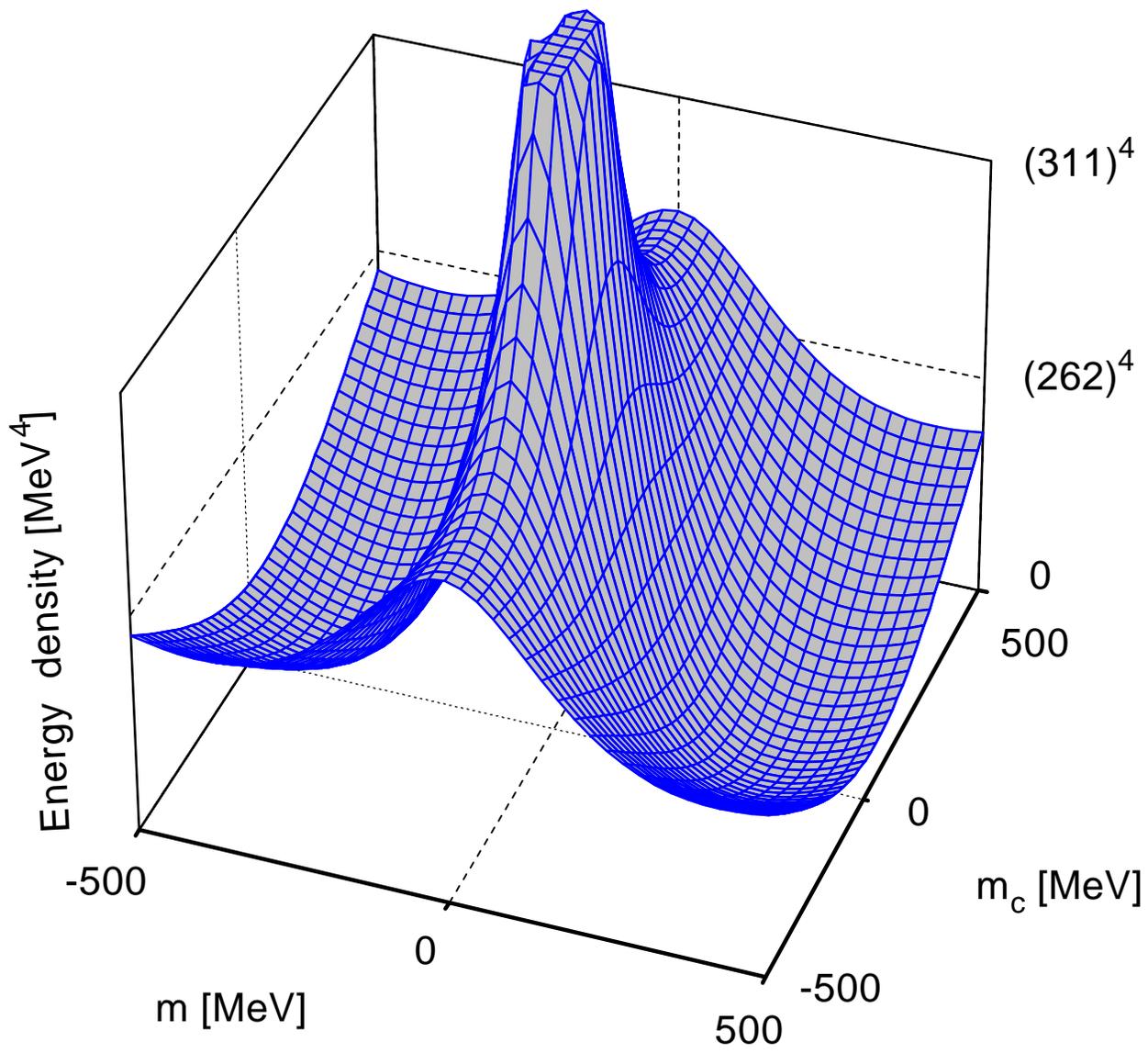
\end{document}